\documentclass{ws-ijmpb}
\usepackage{glossaries}
\usepackage{caption}
\usepackage{xcolor}
\usepackage[verbose]{hyperref}
 \usepackage[autostyle=true]{csquotes}

\hypersetup{colorlinks=false,allbordercolors=blue,pdfborderstyle={/S/U/W 1}}

\begin{document}

\markboth{D. Bounoua, W. Li\`ege, Y. Sidis \& Ph. Bourges}{Orbital currents signature in neutron diffraction}

%
\catchline{}{}{}{}{}
%

\title{Orbital current signature using neutron diffraction}

\author{Dalila Bounoua\footnote{dalila.bounoua@cea.fr}}
\address{Universit\'e Paris-Saclay, CNRS, CEA, Laboratoire L\'eon Brillouin, 91191 Gif-sur-Yvette, France}

\author{William Li\`ege}
\address{Universit\'e Paris-Saclay, CNRS, CEA, Laboratoire L\'eon Brillouin, 91191 Gif-sur-Yvette, France}

\author{Yvan Sidis}
\address{Universit\'e Paris-Saclay, CNRS, CEA, Laboratoire L\'eon Brillouin, 91191 Gif-sur-Yvette, France\\
               Institut Jean Lamour, Universit\'e  Nancy-Artem, Nancy Cedex F- 54011, France}

\author{Philippe Bourges\footnote{philippe.bourges@cea.fr}}
\address{Universit\'e Paris-Saclay, CNRS, CEA, Laboratoire L\'eon Brillouin, 91191 Gif-sur-Yvette, France}

\maketitle


\begin{abstract}
 We review the hallmarks of orbital loop currents in various correlated  electron materials and how  they have been evidenced using polarized neutron diffraction. Over the last 20 years, loop current signatures have been observed in high temperature copper oxide superconductors, iridates, copper oxides spin ladders and recently kagome vanadate superconductors. Such currents induce orbital magnetic moments within the unit cell of these quantum materials that can be detected through their interaction with the neutron spin. In addition to the usual description of orbital moments using point-like local magnetic moments, we here show an alternative description of the neutron magnetic cross-section involving the microscopic currents running between different atomic orbitals. We discuss the corresponding magnetic structure factors and the resulting quantitative differences between both approaches. 
\end{abstract}



\section{Introduction}

Orbital loop currents have been one recent noticeable development to describe novel ground states and phenomena in condensed matter. They have been first discussed in the context of high temperature copper oxide superconductors\cite{Varma06,Varma20,Scheurer18,Sarkar19} where loop currents occur in the prominent but still mysterious pseudogap phase of these materials. Using polarized neutron diffraction, our team played a major role in the rising achievement that the loop currents concept is unavoidable for cuprates\cite{Bourges21} and recent developments\cite{Bounoua22,Bounoua23} may represent the missing link to understand the pseudogap physics.

More recently, their occurrence has also been discussed in a wider range of quantum materials from both experimental and theoretical point of views.  However, one major limitation for the establishment of the loop currents concept in quantum matter is that their experimental observations are usually elusive or hidden, and therefore are regularly challenged resulting in alternative theoretical approaches being more widely considered. This has certainly hindered the broader development of the loop-current concept. 

The orbital loop currents phenomenon is an example of multipolar order in quantum materials. In cuprates, it is associated  with a toroidal moment or anapole polar vector, which appears at the same order as quadrupoles in the expansion of the magnetic energy\cite{diMatteo12}. The anapole  does not exhibit any pole and corresponds to a system of currents that does not radiate into the far field. It is in principle formed by winding a current around a torus. Remarkably, a two-dimensional cut of the torus reveals two loop currents turning clockwise  and anti-clockwise\cite{Bourges21} as for the loop currents order originally proposed in cuprates\cite{Varma06}. In general, loop current would occur in materials with strong interplay between topology, unconventional superconductivity and strong electron–electron correlations, and a wide range of materials could fall into this category.  

Indeed, in addition to high-Tc cuprates where they may play a significant role for the pseudogap physics, the existence of orbital loop currents seems ubiquitous in a wide range of quantum materials. In particular, similar type of orbital loop currents phases have been reported using polarized neutron diffraction in iridates\cite{Jeong17},  spin-ladder materials\cite{Bounoua20} and kagome superconductors\cite{Liege24}. In iridates,  the loop current phase was first report on the hidden order using optical second harmonic generation (SHG)\cite{Zhao16}. 

 These phases are also proposed to be at the origin of the emergence of anomalous physical properties in quantum materials.  This is the case of recent reports suggesting that these phases can be relevant for various transport properties, including  anomalous Hall effect for instance, in materials consisting of transition metal–doped (Bi,Sb)$_2$Te$_3$\cite{Chang13} or  in  twisted bilayer graphene aligned to hexagonal boron nitride\cite{Serlin19}.  In ferrimagnetic Mn$_3$Si$_2$Te$_6$,  an exotic quantum state is driven by ab plane chiral orbital currents flowing along edges of MnTe$_6$ octahedra\cite{loopMn}. The c axis orbital moments induced by the in-plane orbital currents couple with the ferrimagnetic Mn spins, drastically enhancing the ab plane conductivity and yielding colossal magnetoresistance when an external magnetic field is applied along the hard c-axis. In CsV$_3$Sb$_5$, chiral transport was reported in the absence of structural chirality\cite{Guo22}. Remarkably, the transport handedness is found to be switchable by the application of small magnetic fields, and is interpreted as arising from time-reversal symmetry breaking orbital loop currents. Together, these examples highlight the unique magnetoelectric character of loop-current states, in which charge and magnetic degrees of freedom are intrinsically intertwined within a single order parameter, underscoring the broader relevance of the loop-current paradigm for understanding and controlling emergent physical properties in quantum materials such as colossal magnetoresistance or chiral transport.

In this short review, we will first describe the neutron diffraction results in cuprates where the loop current signatures are the most documented.  Next, we will describe how to reformulate  the neutron  cross-sections for such a complex magnetic object, going beyond the simple point-like moment induced by closed loop currents. Finally, we will review neutron diffraction results indicating loop currents in other strongly correlated materials. 

\section{Copper oxide superconductors}

\begin{figure}[tbp]
\includegraphics[width=13 cm]{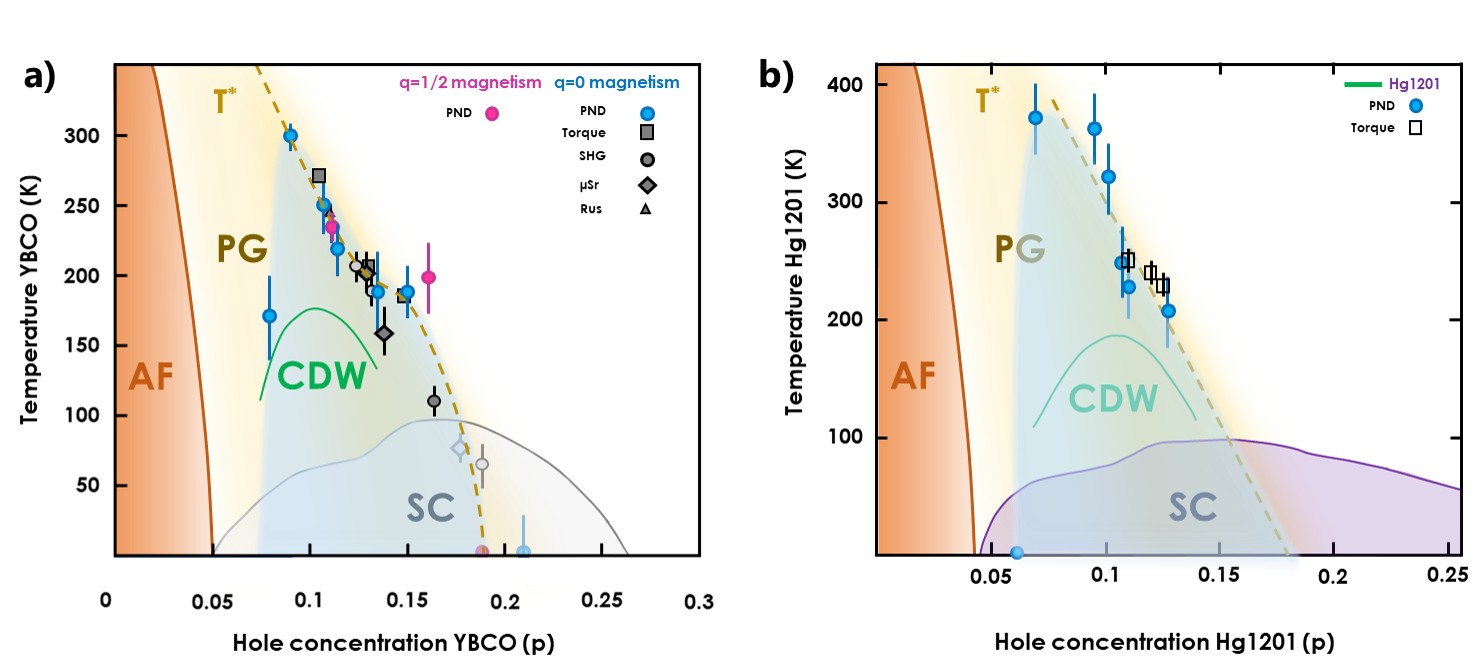}
\caption{{\bf High-$\rm T_C$ superconducting (SC) cuprates phase diagram} versus hole concentration showing broken symmetries at the pseudogap (PG) temperature, T$^*$, that can be accounted for by loop currents:  \textbf{(a)}  for  $\rm YBa_2Cu_3O_{6+x}$ (YBCO) bilayer system and \textbf{(b)}  for the single layer material $\rm HgBa_2CuO_{4+\delta}$ (Hg1201).  The figures are adapted from\protect\cite{Bourges21} where the ${\bf q}=0$ intra unit cell (IUC) magnetism was reviewed along with other techniques which reported broken symmetries at T$^*$  (see text). AF and CDW refer to antiferromagnetism and charge density wave, respectively. Data for the ${\bf q}=1/2$ magnetism in YBCO are included from\protect\cite{Bounoua22,Bounoua23}.  Absence of  ${\bf q}=0$ IUC  magnetism at low doping in Hg1201 has been recently reported\protect\cite{Anderson24}.}
\label{Fig:1}
\end{figure}

 Superconducting cuprate materials are remarkable not only because of their high superconducting critical temperature, but also owing to the occurrence of a mysterious state of matter, the pseudogap (PG) phase, out of which superconductivity emerges\cite{Keimer15}. Below its onset temperature T*, this state is characterized by the disappearance of large portions of the Fermi surface, which reduces to Fermi arcs\cite{Norman98}.  The thermodynamic signature of a phase transition to the pseudogap was first pointed out using high precision magnetization measurements in  $\rm YBa_2Cu_3O_{6+x}$ (YBCO)\cite{Leridon09}. Later, resonant ultra-sound (Rus)\cite{Shekhter13} and subsequently torque measurements\cite{Sato17,Murayama19},  provided compelling evidence that the  pseudogap phase was a true symmetry breaking state. However, the nature of the order parameter and the broken symmetry remain elusive. There are several experimental evidences showing that discrete symmetries are broken at T*, such as the inversion (or parity P) symmetry as reported by SHG\cite{Zhao17},  the time reversal (T) symmetry through multiple polarized neutron diffraction experiments\cite{Bourges21,Bourges11},  Kerr effect\cite{Xia08}, muon spin spectroscopy ($\mu$Sr)\cite{Zhang18,Zhu21},  as well as the fourfold (rotational R) symmetry deduced from torque measurements\cite{Sato17,Murayama19}. Fig. \ref{Fig:1} shows the actual phase diagram versus hole doping for two cuprate families,  YBCO and $\rm HgBa_2CuO_{4+\delta}$ (Hg1201), gathering data points from various techniques reporting a  broken symmetry at T*. The pseudogap vanishes with increasing doping across the cuprates phase diagram (Fig. \ref{Fig:1}). It is usually depicted by  a linear behaviour although it can depart from that as it is suggested here in YBCO.  The discussed  broken symmetries occur exclusively in the mysterious pseudo-gap phase and can be readily understood by the existence of orbital loop currents as we shall discuss in this review.

\subsection{${\bf q}=0$ intra-unit-cell order}

Among the above-mentioned experimental evidences of  discrete symmetry breaking  in the pseudogap state, the polarized neutron diffraction technique played a major role to reveal the existence of a magnetic order, defined as a ${\bf q}=0$ intra-unit-cell (IUC) order. The ${\bf q}=0$ polarized neutron diffraction results have been recently reviewed in ref \cite{Bourges21}.  The observed IUC magnetism is reported in the latest experimental phase diagrams  (Fig. \ref{Fig:1}) and suggests that the IUC order is bound to the pseudogap state, preserving the lattice translational symmetry invariance  (${\bf q}=0$), but breaking discrete Ising-like symmetries: time reversal (T), parity (P) and the fourfold rotation (R) symmetries. This has been established in three different cuprates famillies: YBCO\cite{Fauque06,Mook08}, Hg1201\cite{Li08} and Bi$_2$Sr$_2$CaCu$_2$O$_{8+\delta}$\cite{deAlmeida12}. In a fourth cuprate material, La$_{1.9}$Sr$_{0.1}$CuO$_4$, the magnetic signal is found at short range and  at lower temperature\cite{Baledent10}. The intra–unit-cell order has been as well visualized at low temperature in  Bi$_2$Sr$_2$CaCu$_2$O$_{8+\delta}$  copper oxides using  Fourier transform analysis of spectroscopic imaging scanning tunneling microscopy\cite{Fujita14}. In that later case, the IUC order is detected  through the breaking of the R symmetry. 

Remarkably, in these different cuprates, the IUC magnetism shows up concomitantly with the PG state previously established through many different physical properties\cite{Keimer15}. For instance, in YBCO,  it maps the slight downturn in temperature of the resistivity\cite{Ito93} associated with the pseudogap onset. The polarized neutron diffraction data allow defining a temperature T$_{mag}$\cite{Fauque06} for the onset of the ${\bf q}=0$ IUC magnetic signal, which matches, in a large hole-doping range p= [0.08-0.2] the pseudogap temperature, T*.  However, at lower  doping, the ${\bf q}=0$ magnetism vanishes in both YBCO\cite{Baledent11} and Hg1201\cite{Anderson24} and does not follow the expected doping dependence of the pseudogap, possibly due to the proximity of the long range antiferromagnetic phase associated with copper spins. Clearly, the observed broken symmetries and the ${\bf q}=0$ IUC signatures correspond to what can be expected from an orbital loop current state\cite{Varma06,Varma20,Scheurer18,Sarkar19,Bourges21}.  For instance,  a clear magnetic moment  is observed along the c axis perpendicular to the CuO$_2$ plaquette, even if a noticeable in-plane magnetic component is also detected (see\cite{Bourges21} for different possible explanations). 

\begin{figure}[tbp]
\includegraphics[width=13 cm]{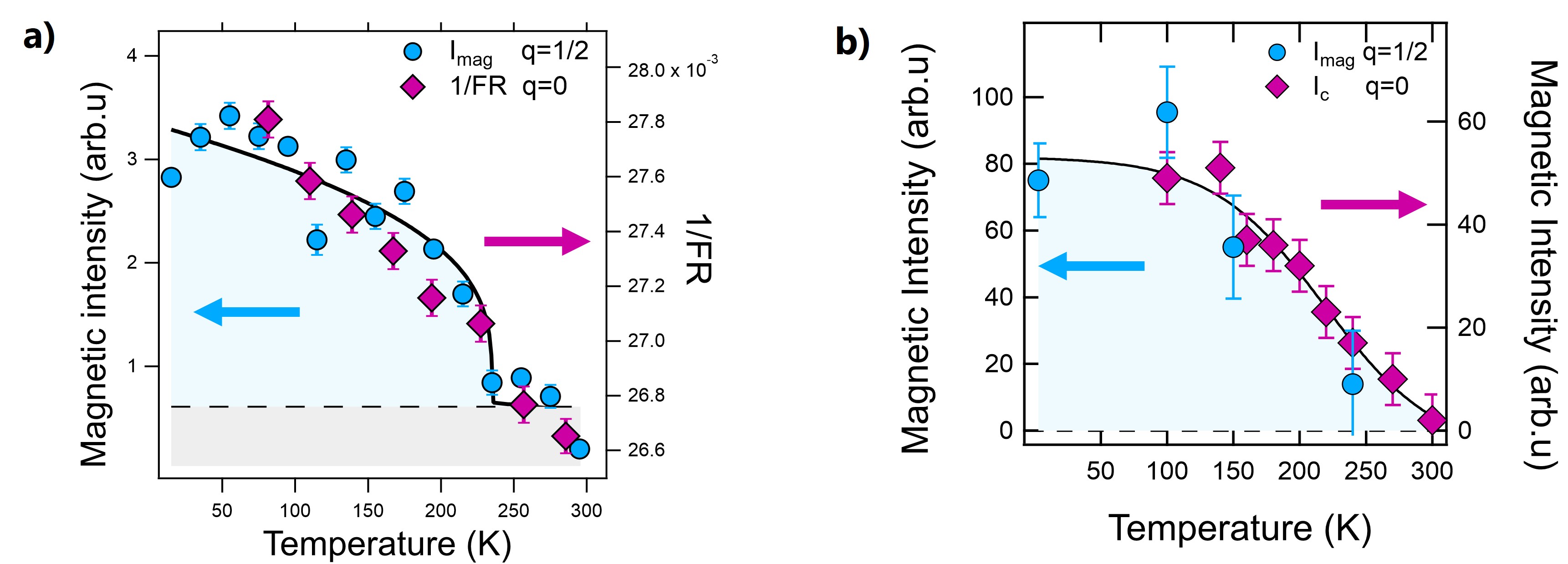}
\caption{{\bf Temperature dependence:}  \textbf{(a)} Comparison of the temperature dependence in the same  YBCO$_{6.6}$  sample of the total magnetic signal at (0.5,0,0) (${\bf q}=1/2$)  as extracted from XYZ polarization analysis (XYZ-PA) (from\protect\cite{Bounoua22}) and the ${\bf q}=0$ IUC  magnetic order measured at the (1,0,0) Bragg peak and represented by the inverse of the flipping ratio  $1/FR$ (from\protect\cite{Mangin17}).  \textbf{(b)}  Comparison of the temperature dependence  in YBCO$_{6.9}$ of the total magnetic signal at (0.5,0,0) (${\bf q}=1/2$)  as extracted from XYZ-PA  (from\protect\cite{Bounoua23}) and the one of the intra unit cell magnetic order (${\bf q}=0$)  measured at the (1,0,0.25) {\bf Q}-position for a similar doping (from\protect\cite{Mangin15}).   }
\label{Fig:2}
\end{figure}

In the bilayer system $ \rm YBa_2Cu_3O_{6.6}$, a strong $a/b$ anisotropy of the intra-unit-cell magnetic structure factor is further observed in detwinned samples\cite{Mangin17}, in addition with a particular $L$ dependence of the structure factor. As a consequence,  the correlation of loop currents between the 2 adjacent  $\rm CuO_2$ plane is actually neither in phase nor out-of-phase, but instead one observes a crisscrossed arrangement of loop currents in the  bilayers\cite{Mangin17}. In terms of anapoles, that could be interpreted as an averaged  bilayer toroidal axis always pointing  along a given direction,  the orthorhombic {\bf b}  direction.

\subsection{Second type of magnetic correlations}

More recently, a second type of magnetic signal has been detected in the pseudogap state of YBCO\cite{Bounoua22,Bounoua23}. This signal was again observed using polarized neutron diffraction, for several hole-doping levels in the same YBCO samples (see Fig. \ref{Fig:1}.a). This second type of magnetism has been detected with a planar propagation wave vector (1/2,0) and/or (0,1/2), defining a ${\bf q}=1/2$ magnetic signal.  While the ${\bf q}=0$ magnetism is at rather long range in the limit of the experimental conditions with 3D magnetic correlation lengths typically larger than $\gtrsim$ 75 \AA \cite{Mook08,Bourges11},  the  ${\bf q}=1/2$  magnetism is at shorter range:  its correlation length has been measured to be  about 6-8 CuO$_2$ unit cells, more precisely  $\xi_{a,b}$ = 24 $\pm$ 4 \AA\ in YBCO$_{6.6}$\cite{Bounoua22}  and $\xi_{a,b} = 30 \pm 3$ \AA\  in YBCO$_{6.9}$\cite{Bounoua23}.  These results suggest short range biaxial magnetic correlations corresponding to a quadrupling (2×2) of the unit cell.

\begin{figure}[tbp]
 \includegraphics[width=13 cm]{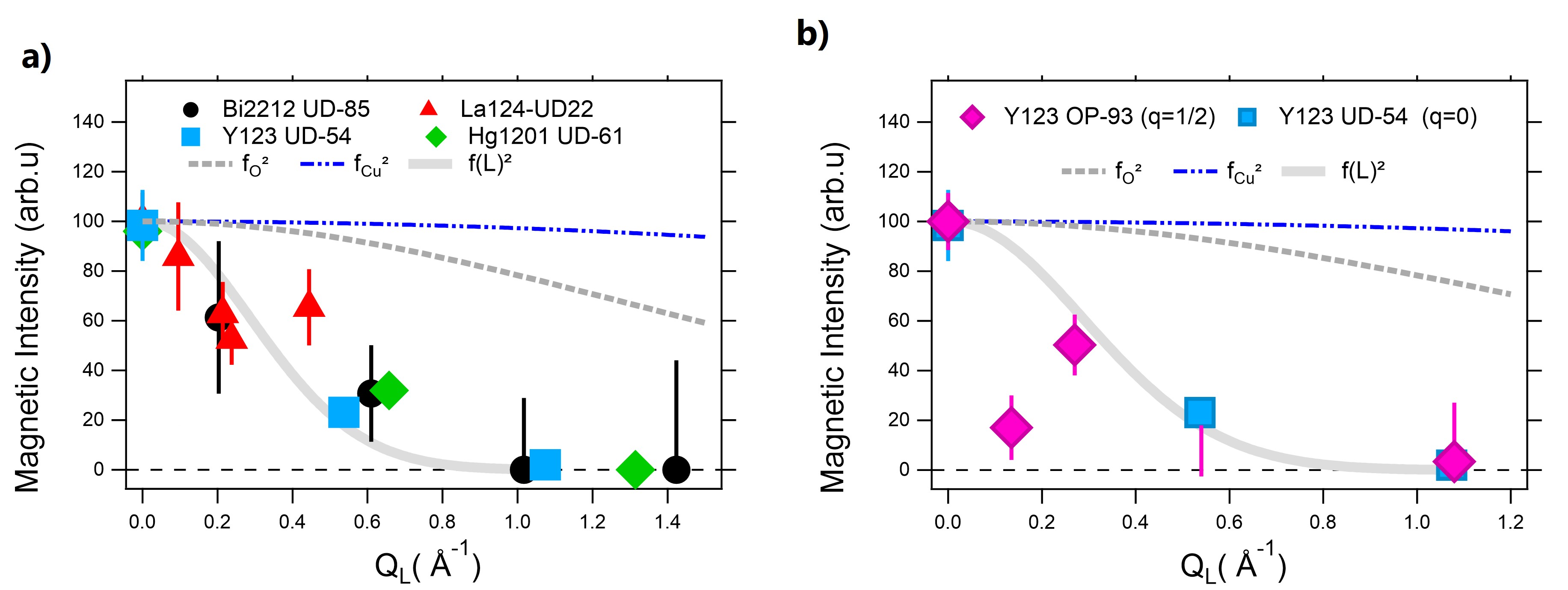}
\caption{{\bf $L$-dependence:}  \textbf{(a)} $L$-dependence of the ${\bf q}=0$ magnetism measured on four cuprate families (from\protect\cite{deAlmeida12}).  The plot is made versus the wavevector $Q_L=\frac{2\pi}{c} L$ in order to compare the four different cuprates. \textbf{(b)} $L$-dependence of the ${\bf q}=1/2$ magnetism measured on two YBCO samples  (from\protect\cite{Bounoua23}).  The low point at $Q_L \sim 0.15$\AA$^{-1}$ is lower due to extra $L$ structure factor dependence. The same form factor is obtained for both signals corresponding to the same grey full line.  }
\label{Fig:3}
\end{figure}

As shown by Fig. \ref{Fig:2}, the ${\bf q}=1/2$ magnetism  exhibits the same temperature dependence as the ${\bf q}=0$ signal for the same doping. This is striking as it was independently measured on different neutron instruments and happens for two different doping levels,  namely, in the underdoped and optimally doped regimes, covering a wide range of doping, while it vanishes beyond the pseudogap state, in the slightly overdoped region.  As we shall see below,  both types of magnetism are related and found to co-exist in the same onset temperature and doping ranges as reported in the YBCO phase diagram (Fig. \ref{Fig:1}.a). The magnetic moment direction has been determined using longitudinal XYZ-polarisation analysis  (XYZ-PA) that allows to measure the different magnetic components.  The largest  magnetic component  at  ${\bf q}=1/2$  is clearly $I_{c}$ associated with  the scattered intensity  along the c-axis,  perpendicular to the CuO$_2$ plane.  Interestingly, XYZ-PA established that only that magnetic component, $I_{c}$, vanishes at $T^{*}$, whereas in contrast,  the much weaker in-plane magnetic component does not change across $T^{*}$\cite{Bounoua22}. At higher doping, similar temperature dependencies of  $I_{c}$ magnetic signals at both wave-vectors have been also reported\cite{Bounoua23}  (Fig. \ref{Fig:2}.b), underlying the close relation between both phenomena. However, at both dopings, the magnetic amplitude of the ${\bf q}=1/2$  signal is much weaker (about 10-15 times) than the ${\bf q}=0$  component.  The short range ${\bf q}=1/2$ magnetism then corresponds to a small volume fraction co-existing with larger ${\bf q}=0$ domains. Interestingly, both types of magnetism exhibit the same magnetic form factor (grey line Fig. \ref{Fig:3}.a-b)  which does not correspond to what would be expected from Cu or even from oxygen spins\cite{Bounoua23,deAlmeida12} but strongly suggesting a common origin.  Finally, let us remark that this second magnetic response at  ${\bf q}=1/2$ has been so far only observed in YBCO. Other cuprates are currently under study to search for the ${\bf q}=1/2$ magnetic signal. 

\subsection{Ruling out Cu spin moments}

We now discuss how to interpret both types of magnetism reviewed so far. First,  let us remind that strong antiferromagnetic (AF) spin correlations are present in copper oxide superconductors in most part of their phase diagram. At low doping, the cuprates phase diagram is characterized by long range antiferromagnetism associated with copper spins\cite{Keimer15}. At higher doping,  spin fluctuations have been observed using numerous techniques among which  inelastic neutron scattering played an important role (see {\it e.g.} Sidis {\it et al}\cite{Sidis04}). For instance, as shown recently  in the single-layer Hg1201 material\cite{Anderson24}, a strong AF response in both the superconducting and pseudogap states is found around  the commensurate wavevector ${\bf q_{AF}}=(0.5,0.5)$ over a wide energy and doping ranges\cite{Chan16}. In the archetypal single-layer cuprate, La(Sr,Ba)CuO$_4$, low energy  magnetic fluctuations are instead found at an incommensurate wavevector shifted from ${\bf q_{AF}}$ 
(see {\it e.g.\cite{Yamada08}}). They are typically associated with copper spin-density-wave or charge stripe correlations, which form magnetic AF domains; the  low energy incommensurate fluctuations have been well documented\cite{Tranquada07} but do not occur in single-layer Hg1201 materials\cite{Anderson24}, which exhibit the undistorted tetragonal prototype structure.

In contrast to the Cu spin low energy fluctuations that remain located at or close to ${\bf q_{AF}}$ with a neutron scattering intensity weighted by the Cu magnetic form factor, the  new magnetic responses discussed above are located either close to ${\bf q}=0$ or ${\bf q}=1/2$ and are characterized by a very unique form factor along $L$ which does not fit with the one of Cu spins.  Likewise, in bilayer cuprates, the low energy response of Cu-spin fluctuations correlate in anti-phase within the CuO$_2$ bilayer\cite{Fong00}, whereas a criss-crossed loop currents arrangement is instead observed for the ${\bf q}=0$ IUC order\cite{Mangin17}.  It is important to stress that neither the ${\bf q}=0$ nor the ${\bf q}=1/2$ magnetic responses can be interpreted using copper spins. Indeed, there is a single copper spin per elementary atomic cell in monolayer materials that cannot produce a ${\bf q}=0$ magnetic response, except in the form of ferromagnetism. This can be ruled out according to macroscopic magnetization  measurements. Further, the ordering at ${\bf q}=1/2$ could be in principle interpreted by spins moments at one site over two Cu-sites with an antiferromagnetic coupling along the unit cell diagonal\cite{Bounoua22}, but  which is not  compatible with the known large first neighbour AF superexchange between copper spins in cuprates\cite{Keimer15,Bourges97}. It is worth stressing that the relation between the AF spin fluctuations response and the ${\bf q}=0$ IUC response, which has been  experimentally established in two cuprates\cite{Chan16}, did not receive theoretical attention, yet.  

\subsection{Orbital loop currents}

In addition to the observed propagation wavevectors which are not consistent with spin ordering, other experimental features favor interpretation of the data in terms of orbital loop currents.  First, a large magnetic moment component is observed to point along the c-axis perpendicular to the CuO$_2$ planes. This is expected for loop currents but is not consistent with spins ordering in cuprates due to their strong XY spin anisotropy\cite{Regnault98}. Second, the observed magnetic form factor alonc c* is unusual with a rapid drop at large L values\cite{Bounoua23,deAlmeida12} (Fig. \ref{Fig:3}), which is inconsistent with the ones expected for spins. Instead, it can support orbital magnetic moments induced by loop currents with an extended geometry in real space along the c axis.  It should be also stressed that the form factor has to  be strongly anisotropic as the fast decay is not observed along the in-plane wavevectors, otherwise no signal would have been detected using polarized neutron diffraction (see next section). 

\begin{figure}[tbp]
	\includegraphics[width=12 cm]{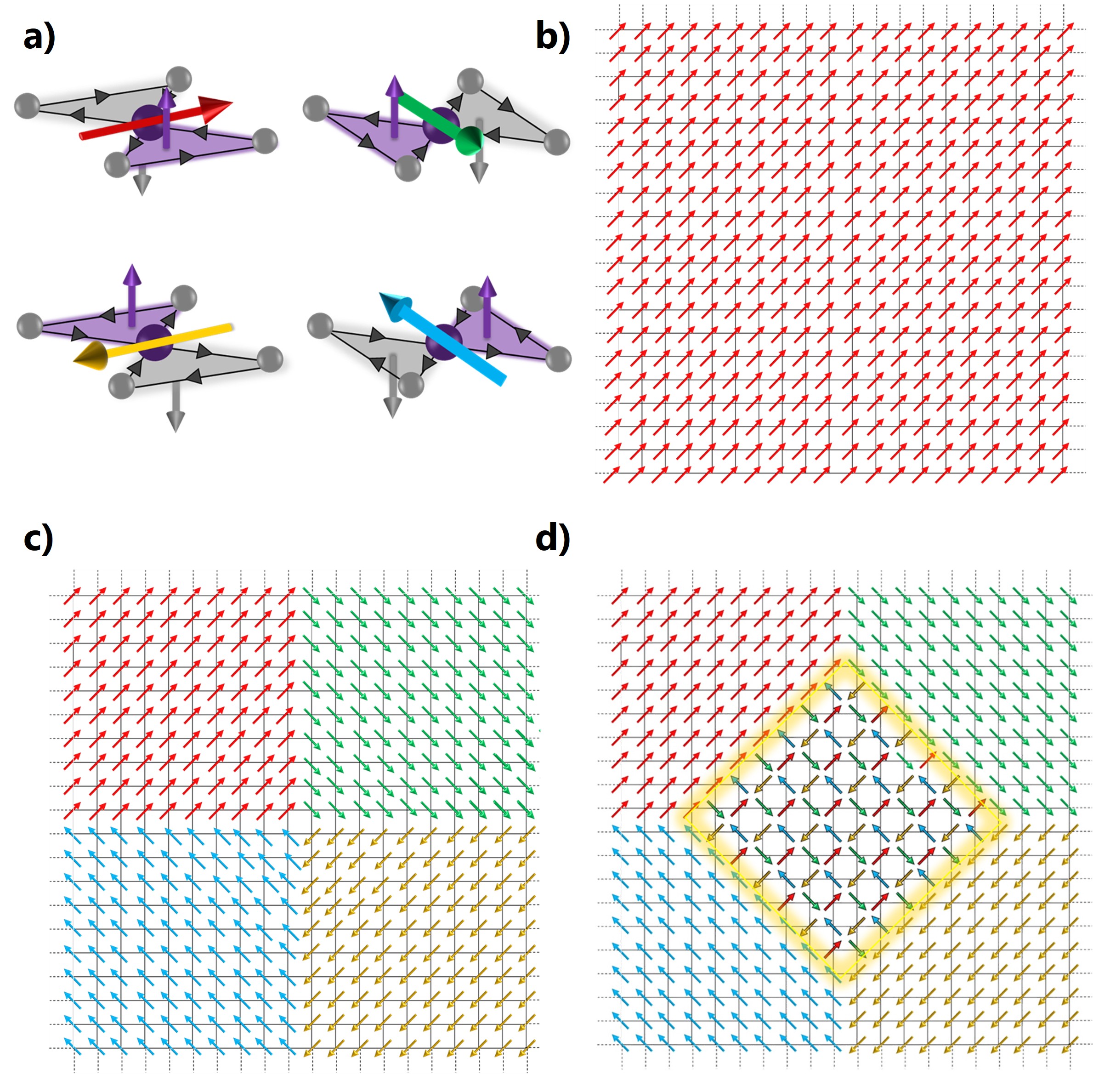}
	\caption{{\bf Orbital loop currents phases}:   \textbf{(a)} Four different degenerated loop currents states. Each spontaneous circulating currents (black arrows) comprises two loops circulating clockwise (in grey) and anti-clockwise (in pruple) leading to two magnetic moments along the c axis  perpendicular to the CuO$_2$ planes. The 4 possible patterns are characterized by horizontal colored arrows, corresponding
		to 4 distinct anapole polar vectors centered at the Cu-site. \textbf{(b)}  Uniform pattern of one anapole state with 20x20 CuO$_2$ unit cells, \textbf{(c)} Anapole vortex pattern of the 4 anapoles as proposed  by Varma\protect\cite{Varma19} and  \textbf{(d)} 2D magnetic texture with 20x20 unit cells paved by the 4 anapoles. The central cluster reproduces the ${\bf q}=1/2$ magnetism with 2x2 loop current patterns binding large ferro-anapolar domains corresponding to the ${\bf q}=0$ magnetism (from\protect\cite{Bounoua22,Bounoua23}).}
	\label{Fig:4}
\end{figure}

Considering that the  ${\bf q}=0$ and  ${\bf q}=1/2$ magnetic responses bear similarities (same temperature dependence with appearance at T*, same form factor, out-of-plane response), one may wonder whether they could share a common origin or not. We proposed a complex magnetic texture\cite{Bounoua22,Bounoua23} where both types of magnetic responses occur in a unified picture. The model is based firstly on the orbital loop current state, referred to as CC-$\Theta_{II}$\cite{Varma06,Varma20}. The loop current phase was first developed as an order parameter for the pseudogap phase. The proposed CC-$\Theta_{II}$ state was actually the primary motivation to search for  ${\bf q}=0$ magnetism in cuprates. It corresponds to a coherently circulating phase between copper and oxygen orbitals within the CuO$_2$ planes,  yielding two opposite orbital magnetic moments perpendicular to it that can be probed with neutron diffraction.  The CC-$\Theta_{II}$ state with {\it m\underline{m}m} symmetry\cite{Varma06} is represented in  Fig. \ref{Fig:4}.a with its fourfold degenerate  ground states. It can be equivalently described by a polar anapole vector either located at the copper site (as in Fig. \ref{Fig:4}.a) or at the center of the unit cell\cite{Bourges21}. The four degenerate ground states correspond to a ${\pi/2}$ rotation of the anapole vector. 

For each of the degenerate ground states, loop currents flowing among copper and oxygen orbitals form a \enquote{butterfly}-shape, which represents the building block of each unit cell. These basic units can correlate in different ways from one unit cell to another leading to different types of orbital magnetic phases. Some examples relevant for interpreting the neutron experiments are shown in (Fig. \ref{Fig:4}.b-d), where only the anapole vector is represented in each unit cell.  In case of a uniform distribution of the anapole vectors  (Fig. \ref{Fig:4}.b), it gives a ${\bf q}=0$  magnetic order (ferro-anapolar order) with a magnetic intensity maximum at  ${\bf Q}$=(1,0,0) or (0,1,0) Bragg peaks as it was experimentally observed\cite{Bourges21,Fauque06,Bourges11}. As the four ground states are degenerate,  four types of domains are possible and contribute to the magnetic Bragg scatterings. 

Interestingly, in case of a coherent correlation between the four domains (as shown in  Fig. \ref{Fig:4}.c, the resulting pattern establishes a new period for the magnetic structure  related to the size of the large domains containing PxP unit cells (P=20 in Fig. \ref{Fig:4}.b-d). It yields loop currents  super-cells, characterized in momentum space by a set of satellites magnetic peaks located at ${\bf \delta}=\pm {1\over{2P}}$ from the nuclear Bragg reflections.  This occurrence was first theoretically proposed by Varma\cite{Varma19} to account for the Fermi arcs observed in the pseudogap state of cuprates\cite{Norman98}. Broken inversion and chiral symmetries were next experimentally reported using circular and linear photogalvanic responses\cite{Lim22} with a deduced chiral and parity-breaking 
 {\it mm2}  symmetry\cite{Lim22} in agreement with the proposed  pattern\cite{Varma19} (Fig. \ref{Fig:4}.c). However, that pattern has not yet been confirmed using diffraction techniques, most likely because the expected ordering incommensurate wavevector is very small. So that, the magnetic satellites remain hidden by the instrumental neutron resolution function in the nuclear Bragg peak shoulders\cite{Bounoua23}. Other diffraction techniques such as resonant X-ray scattering should be employed to observe such satellites although one needs first to establish what would be the signature of loop currents in resonant X-ray experiments.

Following these considerations, we realized that  a loop current model with the smallest possible domain size P=1\cite{Bourges21,Bounoua22}, that can be described as an anapole vortex,  can account for the 2×2 larger unit cells related to the  ${\bf q}=1/2$ ordering. We then developed a complex magnetic texture, represented in Fig. \ref{Fig:4}.d, to account for both long range ${\bf q}=0$ magnetic ordering and the short range  ${\bf q}=1/2$ correlations\cite{Bounoua22,Bounoua23}. That texture mixes smaller regions in the CuO$_2$ plane having loop currents that break the lattice translational symmetry with larger regions in the CuO$_2$ plane that respect the translation symmetry. The size of the 2×2 anapole vortex over the total size of the pattern determines the ratio of intensities between both features assuming that the local orbital magnetic moment amplitude is the same at every site. That complex loop current texture is able to produce a magnetic pattern combining broad scatterings at ${\bf q}=1/2$  and sharper satellites so close to Bragg reflections (1,0)/(0,1) that they cannot be resolved experimentally\cite{Bounoua23}.  This texture of uniform and vortex anapoles then offers a nice description of the neutron structure factor in momentum space.

\section{Polarized neutron cross sections}

The multiple results in cuprates suggest that a loop current state could be a generic feature of their phase diagrams. To learn more about their nature, it is important to be as quantitative as possible. It is then interesting to discuss the neutron scattering cross section associated with loop currents, which is a rather unusual object. In particular, as the building block is the \enquote{butterfly}-pattern, one needs to calculate the structure factor associated to each of the individual patterns shown in Fig. \ref{Fig:4}.a. 

 In case of elastic magnetic scattering, the neutron intensity is proportional to the square of the magnetic structure factor  $\vert F_M({\bf Q}) \vert ^2 $. The magnetic neutron cross section can then  be written as\cite{Squires},

\begin{equation}
{\frac{d \sigma}{d \Omega}}\Big{\vert}_{mag} = \sum_{\tau} \vert F_M({\bf Q}) \vert ^2 \delta({\bf Q} - {\bf \tau})
\label{eq0}
\end{equation}

$F_M({\bf Q})$ is the magnetic structure factor at the momentum tranfer ${\bf Q}$.  ${\bf \tau}$ represents  the propagation wavevector of the loop current state  where the magnetic signal is expected in momentum space depending on the correlations between individual  \enquote{butterflies}. Following our discussion above,  ${\bf \tau}$  corresponds to the Bragg reflections of the lattice for ${\bf q}=0$  or the Brillouin zone boundary for  the  ${\bf q}=1/2$ correlations. The  $\delta$-function in Eq. \ref{eq0} can be replaced by a Gaussian or Lorenztian function in case of short range correlations as it is the case for   ${\bf q}=1/2$ magnetism\cite{Bounoua22,Bounoua23}.

For a polarized neutron diffraction experiment,  the magnetic structure factor reads\cite{Squires},

\begin{equation}
 F_M({\bf Q})  = \frac{r_0}{2} < \pm \vert {\bf \sigma}.{\bf B}({\bf Q}) \vert \pm > 
\label{FM}
\end{equation}

${\bf \sigma}$ are Pauli matrices describing the neutron spin which can take the 
two states: $|+>$ and $|->$. They are related to neutron moment 
as ${\bf \mu_N} =-\gamma\mu_N {\bf \sigma}$  where $\mu_N$ is the nuclear magneton
and $\gamma$=1.913 is the gyromagnetic ratio. The prefactor $r_0 =0.54\ 10^{-12}$ cm corresponds 
to the neutron magnetic scattering length for a magnetic moment of 1 Bohr magneton 
($\mu_B$) ($r_0/\gamma$ is the classical radius of the electron). Using Eq. \ref{FM}, one can take advantage of the great strength of polarized neutron diffraction using XYZ-polarization analysis (XYZ-PA)\cite{Bounoua22,Bourges11} to extract the magnetic signal from the background and determine its different components. 

The key term in Eq. \ref{FM}, ${\bf B}({\bf Q})$, is proportional to the Fourier transform at the 
wave vector {\bf Q} of the magnetic field distribution in real space, ${\bf B({\bf r})}$. There are two possible electronic sources of magnetic field in solids which can interact with the neutron spin, either the magnetic field ${\bf B_S}$ associated with  unpaired electron spins, or the magnetic field  ${\bf B_L}$ related to electron momentum or currents that can generate orbital moments. ${\bf B}({\bf Q})$ is then in general written as a sum of the spin part and the orbital part as\cite{Squires}: 

\begin{equation}
{\bf B}({\bf Q}) = \sum_j \exp(-i  {\bf Q.r}_j) \big\{ {\bf \hat{Q}} \wedge  {\bf M_S} \wedge  {\bf \hat{Q}} 
+ i {I\over{2\mu_B}} \frac{{\bf \hat{Q}} \wedge d{\bf l}_j}{Q} \big\}
\label{Bq-all}
\end{equation}

where ${\bf \hat{Q}}={\bf Q}/Q$.  $\bf M_S$ is the spin moment, $I$ is current intensity along the current path $d {\bf l}$.
The sum is made over all magnetic sites, $j$, within each unit cell. Usually, only the first spin term is considered in neutron experiments.  Here, we only discuss the second one as the copper spins do not contribute for both the ${\bf q}=0$ and  ${\bf q}=1/2$ magnetism.

For the second term, there are two, in principle assumed to be equivalent,  ways to express the neutron cross section arising from the loop currents phase. On the one hand  (see {\it e.g.}\cite{Bounoua22,Bourges11}),  one considers  a point-like orbital moment  that is produced by the circulating current similar to the spin term in Eq. \ref{Bq-all}. On the other hand  (see { \it e.g.}\cite{Sidis07}), ${\bf B}({\bf Q})$ can be expressed in terms of the Fourier transform of the current density. We describe below these two methods to calculate the  \enquote{butterfly} structure factor for the four CC-$\theta_{II}$ states. It should be stressed that all the calculations in the next sections are strictly only valid for the ${\bf q}=0$ order where a given pattern is repeated over all the unit cells. 

For the sake of simplicity in comparing the resulting structure factors, only one  CuO$_2$ layer per unit cell will be considered. However, in bilayer systems, the structure factor contains and additional $L$-dependent factor being related to the correlation of loop currents between the two adjacent  $\rm CuO_2$ planes as it has been reported in $ \rm YBa_2Cu_3O_{6+x}$\cite{Mangin17}  (see above). 

\subsection{ \label{sec:level2A} Orbital point-like moment picture of the CC-$\theta_{II}$ phase}

\begin{figure}[tbp]
\includegraphics[width=12 cm]{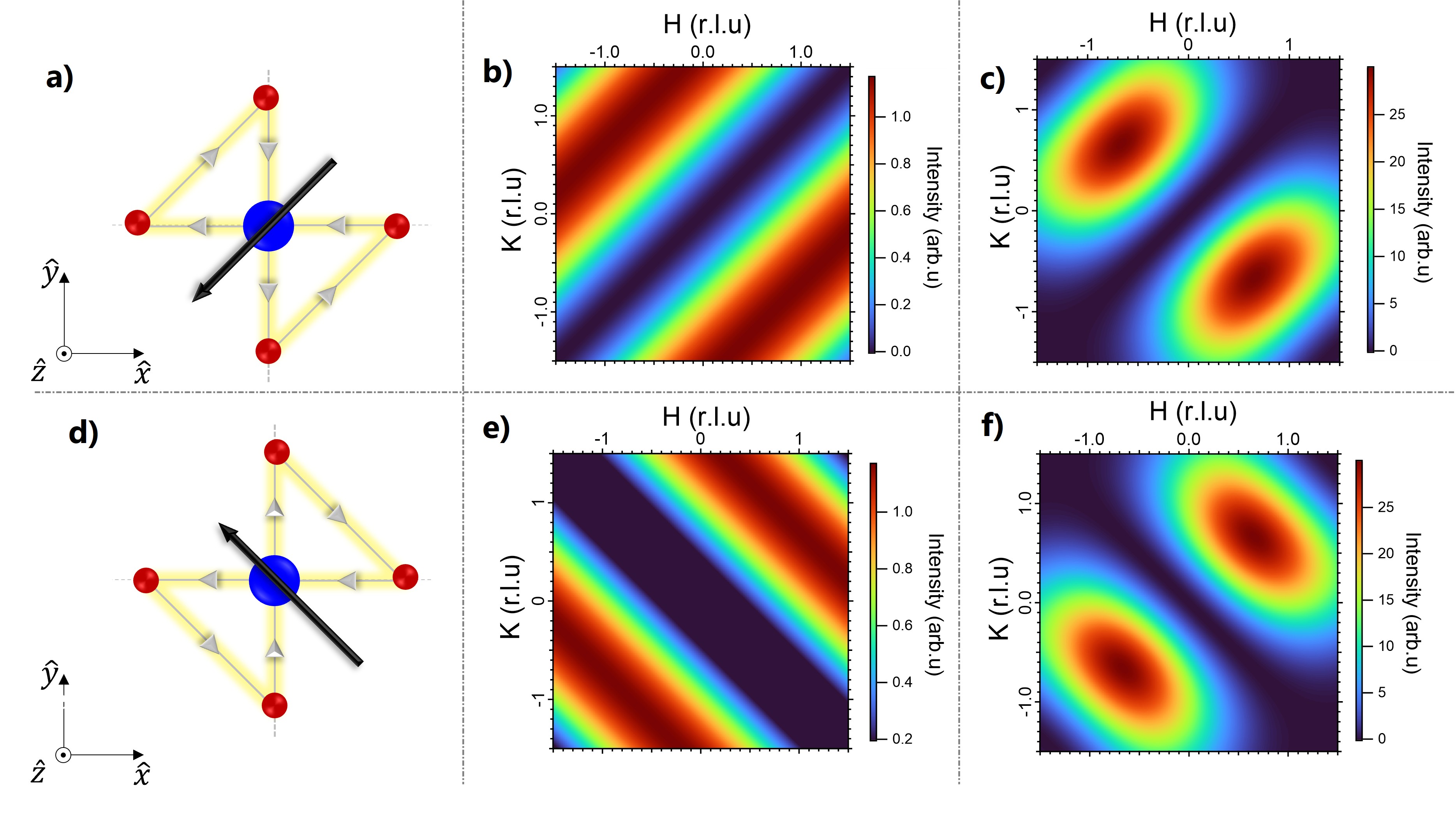}
\caption{{\bf Magnetic structure factors}:   Magnetic moment distribution for two ground states with \textbf{(a)} the anapole along x+y (pattern $\mathcal{P}_a$) and \textbf{(d)} the anapole along  x-y  (pattern $\mathcal{P}_b$).  Magnetic structure factors calculated for a point-like moment located at the center of the loop triangle:  \textbf{(b)} for the anapole along x+y (pattern $\mathcal{P}_a$) and \textbf{(e)} for the anapole along x-y (pattern $\mathcal{P}_b$).   Magnetic structure factors calculated from the currents:  \textbf{(c)} for the anapole along x+y (pattern $\mathcal{P}_a$) and \textbf{(f)} for the anapole along x-y (pattern $\mathcal{P}_b$).  The form factor $f(Q)$ is not included in the calculations.  $(\hat x,\hat y,\hat z)$ defines an orthonormal coordinate system for a tetragonal lattice. }
\label{Fig:5}
\end{figure}

First, one considers, for the orbital part in Eq. \ref{Bq-all}, point-like orbital moments induced by the loop currents. For the CC-$\theta_{II}$ phase, there are two patterns with anapoles along both diagonals:  $\pm$(1,1,0) ($\mathcal{P}_a$) and $\pm$(1,-1,0) ($\mathcal{P}_b$)   directions among the four ground states. The contribution of the two other domains (\enquote{time-reversal domains}) gives the same structure factor as it depends only on the sign of orbital magnetic moment ${\bf M_c}$ pointing along the c-axis.  Both patterns have a different magnetic neutron structure factor.  For the pattern $\mathcal{P}_a$, one needs to put opposite moments at ($-x_0,x_0$) and ($x_0,-x_0$) respectively (Fig. \ref{Fig:5}.a). For the pattern $\mathcal{P}_b$,  the orbitals opposite moments are located at the coordinates   $\pm$($x_0,x_0$) (Fig. \ref{Fig:5}.d) along the unit cell diagonals with  $x_0=0.146$, that defines the geometric center of the loop triangle.  The calculated magnetic interaction ${\bf B}({\bf q})$  can then be written as,

\begin{equation}
{\bf B}({\bf q}) = 2 i\ f(Q)\ \sin (2 \pi x_0 (H \pm K)) ({\bf \hat{Q}} \wedge  {\bf M_c} \wedge  {\bf \hat{Q}})
\label{Bq-M}
\end{equation}

The sign -(+) stands for each pattern, $\mathcal{P}_a$ and $\mathcal{P}_b$ respectively. One needs to associate a magnetic form factor $f(Q)$ ($f(0)=1$) to the orbital magnetic moment ${\bf M_c}$ which should be related to the Fourier transform of involved magnetic atomic orbitals. In principle,  $f(Q)$ should be given by the Fourier transform of the moment distribution associated with a single triangle of the CC-$\theta_{II}$ phase. However, this quantity is not obvious to define but should, in principle, be a combination of copper and oxygen orbital form factors. 

The Fig. \ref{Fig:5}.b and  Fig. \ref{Fig:5}.e  show the magnetic neutron intensity  $|F_M({\bf Q})|^2$  in the (H,K) plane calculated using the point-like model for both patterns. For a sake of comparison, no form factor is included. First, the intensity is zero for H=K=0 due to compensated moments within each unit cell.  Neutron intensity maps are rotated by 90$^\circ$ to each other with the main contribution to the intensity being located along the diagonals.  In Fig. \ref{Fig:6}.a, the  averaged intensity of both patterns assuming equally populated domains is represented, the intensity is maximum along a* and b*\cite{Bourges11}.  The neutron intensity would decay at large Q only due the form factor $|f(Q)|^2$. 

\subsection{\label{sec:level2B} Array of currents of  the CC-$\theta_{II}$ phase}

A second method to treat the orbital term in Eq. \ref{Bq-all} is to consider the internal magnetic field originating from the circulating currents\cite{Sidis07,Hsu91,Chakravarty01}. For a single CuO$_2$  plane, ${\bf B}({\bf Q})$ can be expressed in terms of the Fourier transform, ${\bf j}({\bf Q})$, of the current density ${\bf j} = I  d {\bf l}$ (represented on  Fig. \ref{Fig:5}, for both possible CC-$\theta_{II}$ ground states associated with anapoles at 90$^\circ$). ${\bf B}({\bf Q})$ entering  in  Eq. \ref{FM} is then

\begin{equation}
{\bf B}({\bf Q}) = -i\ f(Q)\ \frac{ {\bf \hat{Q}} \wedge {\bf j}({\bf Q})}{Q}
\label{Bq-jq}
\end{equation}

To calculate for each domain the magnetic scattering signal in the  $Q=(H,K)$ reciprocal space, we need to determine ${\bf j}({\bf Q})$ for a given geometry of microscopic wires. First, one defines the current in real space.  For instance,  for the thread along the $x$ direction in  Fig. \ref{Fig:5}.a, it gives ${\bf j}_1({\bf r})=-I\delta(y)\delta(z){\hat x}$ for $-a/2<x<a/2$ where $I$ is the current flowing in the wires and  $a\simeq 3.85$\AA\ is the lattice parameter. For a \enquote{butterfly} current distribution,  such as Fig. \ref{Fig:5}.a, the total current distribution ${\bf j}({\bf r})=\sum_i{\bf j}_i({\bf r})$ is the sum of currents along the different threads forming both triangles.  One needs next to perform the Fourier transform of the current distribution. For instance, for the $j_1$ example above, it yields ${\bf j}_1 ({\bf Q})= -I a \frac{\sin(\pi H)}{\pi H}{\hat x}$. The Fourier transform of the current distribution, ${\bf j}_a({\bf Q})$ and ${\bf j}_b({\bf Q})$, for both \enquote{butterfly} patterns of  Fig. \ref{Fig:5}, reads
\begin{eqnarray} 
{\bf j}_a({\bf Q})  =& -Ia\frac{\sin(\pi H)}{\pi H}{\hat x} -Ia\frac{\sin(\pi K)}{\pi K}{\hat y} + Ia\frac{\sin(\pi (H+K)/2)}{\pi(H+K)/2}\cos(\pi(H-K)/2)({\hat x}+{\hat y})  \nonumber \\
{\bf j}_b({\bf Q})  =& -Ia\frac{\sin(\pi H)}{\pi H}{\hat x} +Ia\frac{\sin(\pi K)}{\pi K}{\hat y} + Ia\frac{\sin(\pi (H-K)/2)}{\pi(H-K)/2}\cos(\pi(H+K)/2)({\hat x}-{\hat y})
\label{jq}
\end{eqnarray}

\begin{figure}[tbp]
\includegraphics[width=13 cm]{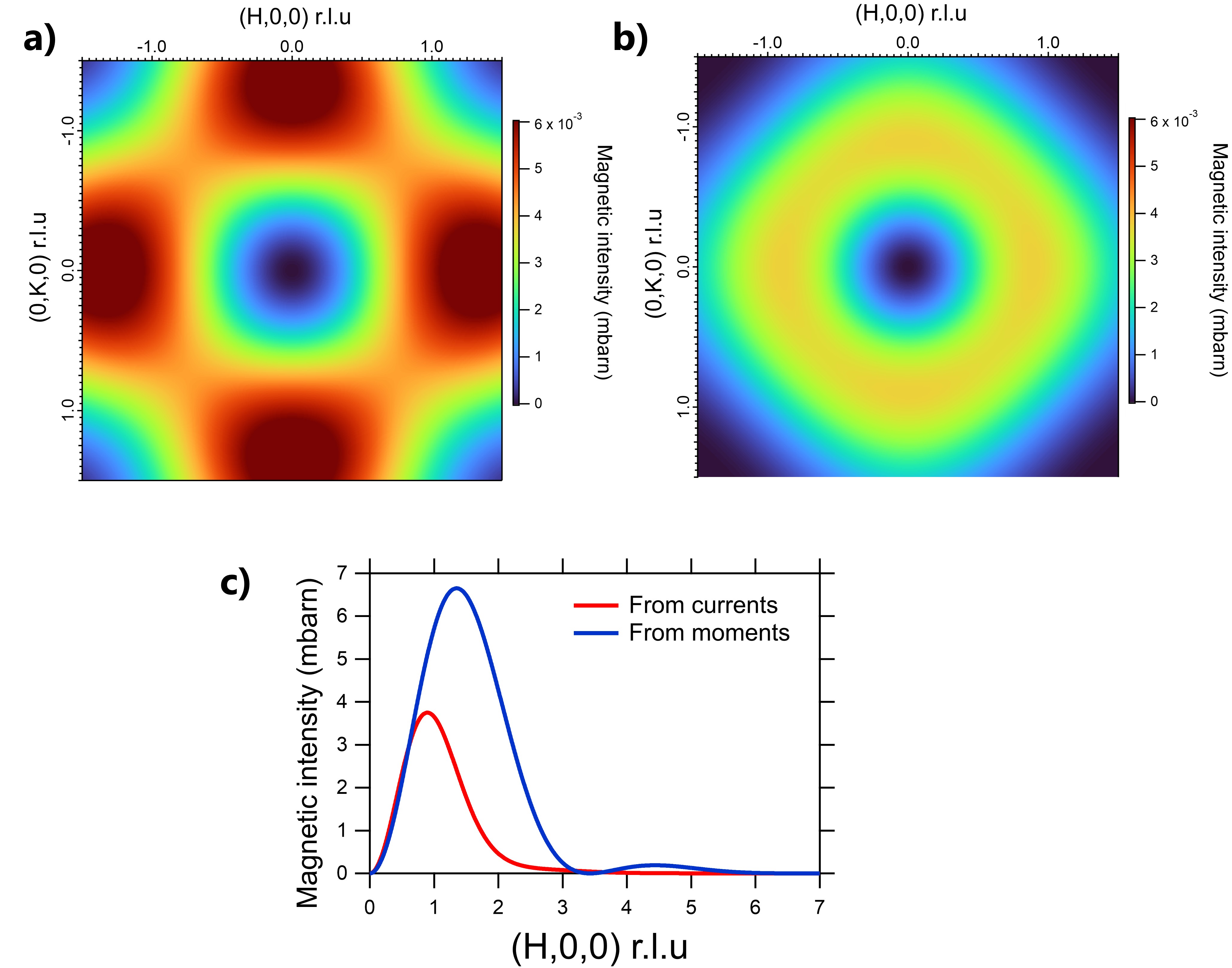}
\caption{{ \bf Structure factor comparison}:  Average intensities of patterns with $\mathcal{P}_a$  and $\mathcal{P}_b$  anapoles as shown in Fig. \ref{Fig:5} for  \textbf{(a)} the point-like model and \textbf{(b)} the currents model.  \textbf{(c)} Comparison  of the structure factors along the H direction for K=0 for both models and assuming the relation $M_c=Ia^{2}/8$. The structure factors are given in mbarns for $M_c=0.1 \mu_B$, typical value found experimentally\protect\cite{Bourges11,Fauque06}.  In  \textbf{(c)}, the copper form factor $|f(Q)|^2$ is also  included to compare with neutron diffraction experiments.}
\label{Fig:6}
\end{figure}

The neutron structure factor Eq. \ref{FM} can be deduced using Eqs. \ref{jq} in  Eq. \ref{Bq-jq}.
The Fig. \ref{Fig:5}.c and  Fig. \ref{Fig:5}.f  show the magnetic neutron intensity  $|F_M({\bf Q})|^2$  in the (H,K) plane calculated using the currents model for both patterns. Similarities  with the point-like model are clear with a minimum along one diagonal and zero intensity for H=K=0. However,  in contrast with the point-like model, the neutron intensity decays at large Q, even in the absence of the form factor,  due to the 1/Q factor in Eq.  \ref{Bq-jq}.  The averaged intensity of both patterns assuming that they have the same population for both types of domains is represented in Fig. \ref{Fig:6}.b. Being more isotropic, the obtained intensity differs noticeably from the one calculated with the point-like moment model shown in Fig. \ref{Fig:6}.a.

 In this version of the loop current model, the wires are assumed to be infinitely thin. However, in principle,  one  needs to add an effective \enquote{transverse} thickness, $\delta$, of each wire, to account for the typical size of copper and oxygen $d$ and $p$ orbitals. This leads to an additional (still empirical) form factor $f(Q)$ with $f(0)$=1,  which is similar to the magnetic form factor for point-like moments. It is interesting to notice that the two \enquote{transverse} current thicknesses are decoupled from each other and can take distinct values unlike  atomic orbitals for point-like moment.  Therefore, the related form factor can be in principle anisotropic between in-plane and out-of-plane directions. It is worth to remind that this anisotropy is actually observed  experimentally\cite{Bounoua23,deAlmeida12} (see Fig. \ref{Fig:3}) although the exact explanation is not yet established.

Basically, in comparison to spin moments, it is usually considered\cite{Sidis07,Chakravarty01} that the neutron scattering intensities qualitatively differ for orbital currents due to the characteristic size of the current loop which can be as large as  the unit cell distance between the copper atoms, a=3.85\AA, whereas it is the size of the copper orbitals, a$_0 \sim 1.2$\AA, for spins. Therefore, it is generally assumed that the neutron intensity from orbital currents is likely to fall off more rapidly with increasing $Q$ than the usual atomic form factor for a spin moment, that was particularly the case for $D$-density wave model\cite{Hsu91,Chakravarty01}. The effect can be so important that the expected intensity is often considered to be too small to be detected in neutron experiments. The decreasing intensity is indeed observed for currents in  Fig. \ref{Fig:5}. However, because the discussed loop triangle of the CC-$\theta_{II}$ phase corresponds to only 1/8 of the unit cell area. The expected intensity is actually still sizable in cuprates to be measured in neutron diffraction experiments. 

\subsection{ \label{sec:level2C} Comparaison of neutron intensity obtained with both methods}

To be more concrete and compare quantitatively  the structure factor obained via both methods, we expand the relations (Eqs. \ref{Bq-M} and \ref{jq}) for a scattering 
plane defined by the two reciprocal directions $({\bf a*},{\bf c*})$, then ${\bf Q}= 2 \pi H/a {\bf\hat x} + 2 \pi L/c {\bf\hat z}$. This reciprocal plane $(H,0,L)$ was often used as the scattering plane in neutron diffraction experiments.

For the point-like moment model (section \ref{sec:level2A}), one obtains a moment perpendicular to the scattering plane, ${\bf M}=M_c {\bf\hat z}$. In principle, $M_c=I A$ where $I$ is the current density in the wire defined above and $A=a^2/8$ the cross-section area of each triangle. Then, ${\bf \hat{Q}} \wedge  {\bf M} \wedge  {\bf \hat{Q}}$ entering in Eq.~\ref{Bq-M} becomes $ \frac{M_c  H}{H^2+(a/c)^2 L^2} ( -{a\over c} L {\bf\hat x} + H  {\bf\hat z})$. 
Eq. \ref{Bq-M} can be rewritten for both domains as, 

\begin{equation}
{\bf B}({\bf Q}) = \pm 2 i  f(Q)\ \sin (2 \pi x_0 H) \frac{H M_c}{H^2+(a/c)^2 L^2} \big[ -{a\over c} L {\bf \hat x} + H  {\bf\hat z} \big]
\label{Bq-M-H0L}
\end{equation}

$c$ is the c-axis lattice parameter.  For the currents model (section \ref{sec:level2B}) and $K=0$, the current density for both patterns becomes
 ${\bf j}_a({\bf Q}) = -{\bf j}_b({\bf Q}) = Ia [1- \frac{\sin(\pi H)}{\pi H}]  {\bf\hat y}$ and ${\bf B}({\bf Q})$ reads

\begin{equation}
{\bf B}({\bf Q}) =\pm  i  f(Q)\ {{1}\over {2 \pi}} (1- \frac{\sin(\pi H)}{\pi H}) \frac{{Ia^2}}{H^2+(a/c)^2 L^2}  \big[ -{a\over c} L {\bf\hat x} + H  {\bf\hat z} \big]
\label{Bq-H0L}
\end{equation}

Interestingly, Eqs. \ref{Bq-M-H0L} and  \ref{Bq-H0L} show clear common similarities about the neutron  orientation factor. However, 
their in-plane dependencies noticeably differ: the Fig. \ref{Fig:6}.c shows  along $H$ and for $L=0$ both structure factors $|{\bf B}({\bf Q})|^2$ for a single one \enquote{butterfly}-shape loop currents pattern including the form factor $f(Q)$ for  Cu$^{2+}$ spins. Clearly, both methods give different results. To compare both models in absolute units, we use the relation $M_c=Ia^{2}/8$ between the magnetic flux, the current and the area of one triangle $a^2/8$. At H=1, the intensity ratio of both methods is $\sim$ $0.8^2$ = 0.64.  The Fig. \ref{Fig:6}.c shows the possible correspondence with experiments although there is so far no existing possible comparison as no intra-unit-cell magnetism has been yet detected at larger $H$.  For instance, at Q=(1,0,0), the expected magnetic intensity in absolute units is found to be  $\sim 5.7$ mbarn from the point-like-moment magnetic structure factor Eq.\ref{Bq-M-H0L} and  $\sim 3.6$ mbarn from the current calculations of the magnetic structure factor Eq. \ref{Bq-H0L}. These amplitudes are consistent with the reported IUC magnetic intensity at (1,0,0) and (0,1,0) of $\sim 6$ and $\sim 2$ mbarn\cite{Mangin17} as specified in\cite{Bounoua22}.

\section{Other correlated materials}

Loop currents are also discussed as possible instabilities in other quantum materials aside from the superconducting two-dimensional (2D) cuprates. From our perspective, that was not obvious at first as the proposed theory was developed specifically for 2D CuO$_2$ layers\cite{Varma06}. However, in recent years, more theoretical approaches for loop currents as well as novel experimental evidences have been made in other materials. We here shall review these novel developments in the following. 

\subsection {Two-leg ladder cuprates}

The first example does not stray too far from 2D cuprates. The two-leg ladder cuprate Sr$_{14-x}$Ca$_x$Cu$_{24}$O$_{41}$ hosts a very rich phase diagram with a well-established spin liquid state for a wide range of  Ca content. Its crystal structure, called composite, consists in two interacting sub-lattices, $CuO$ chains and $Cu_2O_3$ ladders, which are incommensurate each other along the c axis. 

\begin{figure}[tbp]
\includegraphics[width=12 cm]{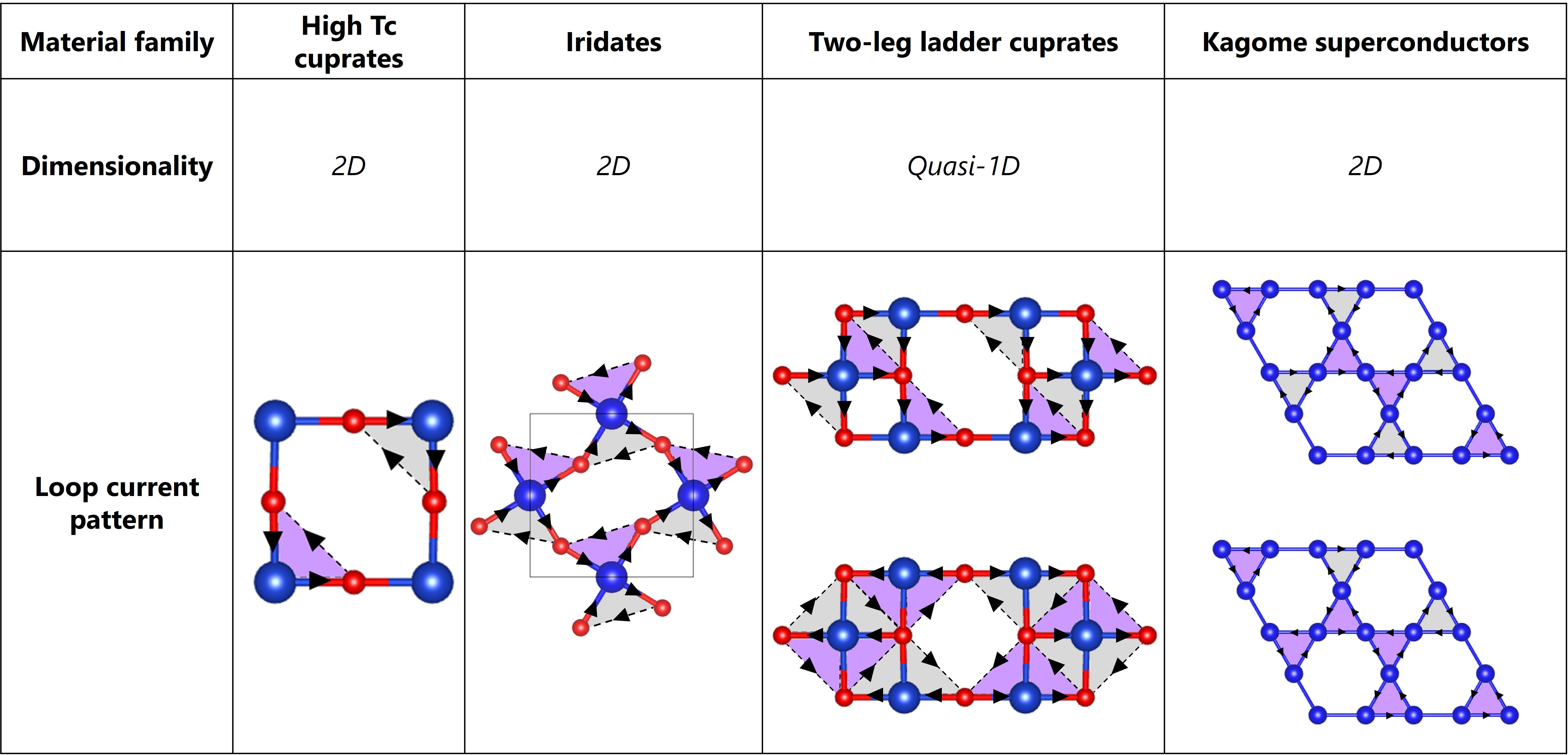}
\caption{{\bf Classification of loop current models}: Different loop current patterns able to reproduce the  ${\bf q}=0$ magnetism in copper oxide superconductors including YBa$_2$Cu$_3$O$_{6+x}$ and HgBa$_2$Cu$_4$0$_{4+x}$\protect\cite{Bourges21}, iridates Sr$_2$(Ir,Rh)O$_4$\protect\cite{Jeong17,Zhao16} and quasi-1D two leg ladder cuprates Sr$_{14-x}$Ca$_x$Cu$_{24}$O$_{41}$\protect\cite{Bounoua20}, the upper model corresponds to the CC-$\theta_{II}$ phase\protect\cite{Varma06} while the lower pattern corresponds to the model proposed in the context of two-dimensional spin-liquids\protect\cite{Scheurer18}. In the metallic kagome CsV$_3$Sb$_5$, two patterns consistent with the neutron diffraction experiments\protect\cite{Liege24} with loop currents on vanadium triangles that respect the CDW  symmetry.}
\label{Fig:7}
\end{figure}

Using polarized neutron diffraction, we observe novel short range magnetic correlations for two different Ca  contents\cite{Bounoua20}.   The magnetic correlations develop exclusively at forbidden Bragg positions of  the two-leg ladders sub-lattice. Models involving magnetic moments on Cu or oxygen sites fail to account for the experimental results whereas simple models based on loop currents on the ladders sub-lattice capture the most salient observation of the magnetic scattered intensity\cite{Bounoua20}. The similarities of the observed phenomena with 2D cuprates are striking.  The neutron data show that 50\% of the magnetic moment lies out of the ladder planes with a tilt of the moment in agreement with previous estimates in superconducting cuprates\cite{Fauque06,Mook08,Tang18}. The short range magnetism remains confined to a single two-leg ladder at low substitution similarly to what has been reported in lightly doped (La,Sr)$_2$CuO$_4$ cuprate where the short-range loop currents correlations are short range, 2D and confined in  bond centered charge stripes\cite{Baledent10}. Of course, the cause for confinement is different in both systems: in the later case, it is due to the doping and spontaneous stripe formation and in the former case due to the intrinsic structure of the material. The deduced loop current patterns, shown in Fig. \ref{Fig:7},  adapt themselves to the ladder quasi-1D atomic structure with increasing inter-ladders correlations upon increasing  the Ca substitution\cite{Bounoua20}.

\subsection{Iridates}

A second example concerns the layered perovskite iridate, Sr$_2$(Ir,Rh)O$_4$, which exhibits strong similarities with high-Tc cuprates in terms of layered peroskite structure and electronic states, in spite of a different nature of 5d (iridium) and 3d (copper) orbitals. At low doping, long range antiferromagnetic ordering is present in iridates suggesting complex spin-orbital state represented by an effective total angular momentum, $J=1/2$\cite{Kim09}. These materials also exhibit a PG-like behaviour reminiscent of cuprates in their electronic excitation spectrum\cite{Kim16}. The motivation to search for loop currents with polarized neutron diffraction in this material was the precursory observation of an hidden order breaking inversion symmetry by SHG\cite{Zhao16} that has been subsequently observed in cuprates\cite{Zhao17}.  Using polarized neutron diffraction,  time-reversal broken symmetry was reported concomitantly with  the hidden order phase\cite{Jeong17} with an onset temperature that matches the odd-parity hidden order reported by SHG experiments upon Rh substitution\cite{Zhao16}. Similarly to loop currents in cuprates  the lattice translation invariance was preserved, suggesting that the novel magnetic order and broken symmetries can be explained by the loop-current model, previously predicted for copper oxide superconductors\cite{Varma06}. That result suggests a more generic character of loop currents in correlated electron oxides. The planar structure that involves oxygen ligand could be a key ingredient for the establishment of the loop current phases as shown Fig. \ref{Fig:7}.

In relation with the AF spin order, there is however an interesting difference between the observation of loop current states in cuprates and iridates that might suggest a different microscopic model for loop current in each material. In iridates, the inversion and time-reversal symmetry breaking are already observed in insulating Sr$_2$IrO$_4$, coexisting with the Néel spin order whereas in cuprates the  ${\bf q}=0$ IUC magnetic ordering vanishes at low doping\cite{Baledent11,Anderson24} (see Fig. \ref{Fig:1} above). The existence of the same loop current states in both insulators and metals has been discussed in the presence of a $Z_2$ topological order\cite{Chatterjee17}. Within this framework, on the square lattice antiferromagnet, loop currents can co-exist with an AF ordered phase with collinear spin correlations as we have reported in iridates. Why it is then not happening in cuprates ? This could correspond to two different limits of the same model, which could explain the difference between cuprates and iridates\cite{Chatterjee17b}.

\subsection{kagome metals}

Another case for loop currents phase recently emerges in metallic  kagome  materials from their nontrivial band structure. A chiral flux phase has been theoretically predicted in the topological superconductor AV$_3$Sb$_5$  (A = K,Rb,Cs)  with a quasi-2D kagome lattice\cite{Feng21,Lin21,Zhou22}. These materials exhibit a topological bond or charge density wave (CDW) as described experimentally using X-ray at the {\bf M}-point of the Brillouin zone\cite{Li21}.  In particular, the kagome metal, CsV$_3$Sb$_5$, exhibits both a CDW below 94K with the 2x2 doubling of the hexagonal unit cell and a superconducting phase below 2.5K. These materials also show strong anomalous Hall effect\cite{Yu21}, but no spin ordering has been found in these materials both by muon spin spectroscopy and neutron diffraction. Instead, the proposed chiral flux state\cite{Feng21,Lin21,Zhou22}  breaks time-reversal symmetry yielding anomalous Hall effect  as it was long-sought-after by Haldane\cite{Haldane88} without relying on spins. Rapidly, time-reversal symmetry breaking were reported in the charge ordered state using muon spin spectroscopy\cite{Mielke22}. That was interpreted by loop current states with a current pattern among the triangles of vanadium atoms as shown in Fig. \ref{Fig:7} but the exact loop currents pattern can only be deduced from diffraction techniques.  Scanning tunnelling microscopy further provides evidence for time-reversal symmetry-breaking  in the superconducting state of  Cs(V,Ta)$_3$Sb$_5$ through the interference between internal magnetism, Bogoliubov quasi-particles and pairing modulation\cite{Deng24}.

We recently perform polarized neutron diffraction on the CsV$_3$Sb$_5$ material  searching for magnetic intensity at momenta positions relative to the CDW\cite{Liege24}. This measurement was experimentally challenging and went close to the limit in accuracy obtainable with polarized neutron diffraction in a reasonable time. Most models predict loop currents to produce magnetic intensity at the Brillouin zone boundary ${\bf M}$-points: ${\bf M_1}$=(1/2,0,L) or ${\bf M_2}$=(1/2,1/2,L) reciprocal space positions with L={0 or 1/2}. We investigated both momenta positions.  For the first one, no magnetic signal was observed ruling out the possibility of having a magnetic moment larger than 0.01  $\mu_B$ per vanadium atom. However, measurements at ${\bf M_2}$ suggest the possibility of a magnetic signal, with a tiny magnetic moment of only  m = 0.02 $\pm$ 0.01 $\mu_B$ per vanadium triangle\cite{Liege24}. This shows that current models have to be refined whether toward a lowering of the expected magnetic moment or toward loop current patterns (such as the ones shown in Fig. \ref{Fig:7})  that give rise to magnetic intensity at reciprocal space positions compatible with the neutron measurements. 

\section{Conclusions and other developements}

Loop currents have emerged over the last decades as a new platform to investigate the interplay between topology, unconventional superconductivity and strong electron–electron correlations. These state corresponds to extended magnetization of the metallic atoms on the ligand atoms through their orbitals hybridization.  Orbital loop currents have been suggested either by various novel theoretical predictions or experimental reports showing unexpected symmetry breakings.  The former happened in kagome metals where numbered theoretical models encourage experimentalists to push the limit of their experimental techniques.  The latter occurred for iridates.
Generalization of loop currents phase in other class of quantum materials would certainly grow in the coming years although their identification is usually elusive experimentally.

In cuprates, the observation of loop currents using polarized neutron scattering\cite{Bourges21} has given a clear ground for an unforeseen solution for the origin of the pseudogap and, as a result, may yield a possible novel scenario for high-$T_c$ superconductivity.  Although uniformly ordered flux phases cannot alone be responsible for the pseudogap,  topological orders\cite{Scheurer18}, bond-density wave\cite{Sarkar19} or pair density wave\cite{Agterberg15}, to which loop currents are inseparable, would be capable to open a gap at the Fermi surface. Further, even with uniform flux phases, non uniform bond currents can occur\cite{Liu26} and can be responsible for a fermionic gap. A loop current superstructure has been as well proposed to explain the pseudogap\cite{Varma19}. This is so far not observed directly through diffraction techniques.  

Experimentally, one instead observes additional short range correlations indicating a complex organisation of anapoles - associated with loop currents - between different unit cells\cite{Bounoua22,Bounoua23}. The ${\bf q}=1/2$  magnetic signal, although very weak, shows a breaking of the translational symmetry with a doubling of the unit cell along both a* and b* directions that could play an important role for the pseudogap physics  together with the ${\bf q}=0$ magnetism. Aiming at assessing the relationship between the pseudogap phase and both magnetic responses, we are currently extending our  study to an underdoped YBa$_2$Cu$_3$O$_{6.4}$ (p=0.07) sample closer to the point where spin antiferromagnetic fluctuations become  prominent.

 Another candidate where to find loop currents is the Fe-based superconducting materials that could also host orbital loop currents\cite{Kang11,Klug18} owing to the spin-orbit coupling present in these compounds, which would enforce the emergence of such an order in addition to the usual stripe-type AF state. The FeSe system is certainly one of the first Fe-based superconductor to look at as there is a nematic state below 90 K and no long-range AF spin order at low temperature.

On a broader view, the loop currents state belongs to the class of multipolar phenomena, a field that has seen an important development in recent years.  For instance, the existence of anapoles in a very different system, the ferromagnetic compound Sm$_{0.976}$Gd$_{0.024}$Al$_2$, was revealed very recently by polarized neutron diffraction\cite{Lovesey19}. Similarly, an anapolar contribution to the magnetism of Cr-doped V$_2$O$_3$ \cite{Lovesey07} was measured using resonant X-ray diffraction. This opens potential new routes for the observation of multipolar physics. More recently, a multipolar order underlying spin magnetism was also predicted in altermagnets\cite{Bhowal24,McClarty24}, for instance in  MnF$_2$ due to the symmetry of the lattice near the Mn site related to neighboring fluorine atoms. These multipoles are thought to be essential to establish altermagnetism. Experiments are currently underway to determine if the predicted magnetic octupoles are relevant or not in MnF$_2$.

 URu$_2$Si$_2$ is another emblematic material with strong electronic correlations and puzzling physical properties, in particular the existence of an enigmatic hidden order below $\sim$ 17.5K.  Recent theoretical attempts describe its complex physics by related orbital current states\cite{Silva20,Dmitrienko18}. A chiral spin liquid has been proposed to account for the hidden order\cite{Silva20}. This state involves different types of loop currents (which do not necessarily break the translation of the lattice). Another approach proposes toroidal quadrupoles, implying a complex magnetic pattern with ferro-vortex and antiferro-vortex moments\cite{Dmitrienko18}.

Finally, loop current physics would certainly be of paramount importance in compounds where quantum criticality and quantum fluctuations play a major role. Cuprates, heavy fermions and iron-based superconductors can be considered as promising candidates. Original ideas induced by exchanges between theory and experiment shall generate new perspectives and the design of novel experiments, in order to firmly establish the loop current paradigm in solid-state science.

\section*{Acknowledgements}
We wish to thank Vivek Aji and Beno\^it Fauqu\'e with who we initiated the currents approach for the neutron diffraction structure factor. We acknowledge financial support from the ANR JCJC project NEXUS (Contract ANR-23-CE30-0007-01) of the Agence Nationale de la Recherche (ANR) French agency and the 2FDN - Fédération Française de la Diffusion Neutronique.

\section*{ORCID}

\noindent Dalila Bounoua  - \url{https://orcid.org/0000-0003-0860-1321}

\noindent W. Li\`ege -  \url{https://orcid.org/0009-0001-9126-2318}

\noindent  Yvan Sidis -  \url{https://orcid.org/0000-0002-9333-0169}

\noindent  Philippe Bourges - \url{https://orcid.org/0000-0001-9494-0789}

\end{document}